# EMPLOYING VIRTUALIZATION FOR INFORMATION TECHNOLOGY EDUCATION

___

Timur Z. Mirzoev, Georgia Southern University

## Abstract

This manuscript presents teaching and curriculum design for Information Technology classes. Today, students demand hands-on activities for the newest technologies. It is feasible to satisfy this appetite for exciting education by employing server virtualization technologies to teach advanced concepts with extensive hands-on assignments. Through utilization of virtualized servers, students are able to deploy, secure and manage virtual machines and networks in a contained environment. Various techniques, assessment tools and experiences will be analyzed and presented by this manuscript. Previous teaching cases for Information Systems or Information Technology classes are done using non-commercial products, such as free VMware Server or VMware Player. Such products have very limited functionality in terms of networking, storage and resource management. Several advanced datacenter functions, such as Distributed Power Management (DPM), vMotion and others, are not available in desktop versions of that type of virtualization software. This manuscript introduces the utilization of commercial software, such as vSphere 4.1, with full datacenter functionality and operations for teaching Information Technology classes of various levels.

## Introduction

Virtualization is utilized by various areas of Information Technology (IT) fields and provides the basis for technologies, such as cloud computing, green computing, server consolidation, disaster recovery and many others [1]. Virtualization has been used in various areas of IT in commercial/business settings [2]. IT infrastructure is rapidly adapting to new technologies based on virtualization. While there is still confusion about cloud computing technology [3], it is one of the innovative IT concepts that is based on server and storage virtualization [4].

Higher education is responding to the advantages of virtualization technologies as well [4-7]. Many schools have approved teaching virtualization courses to students and are now faced with the challenge of providing hardware/software support for even freshmen-level classes [7]. Lungsford's approach, as well as many other authors', to teaching IS or IT classes using virtualization is done using non-commercial products, such as free VMware Server or VMware Player. Such products have very limited functionality in terms of networking, storage and resource management and are not used in commercial datacenters. This manuscript introduces the utilization of commercial software, such as vSphere 4.1, which is also a foundation for cloud computing applications.

When utilizing lab computers for deployment of several virtual machines, there is always a concern for protection of the operating systems of lab computers—students may damage software, which results in the reinstallation of operating systems and software packages. A typical solution to "secure" a lab environment is to lock computers so students are unable to change vital setting on laboratory machines [7]. However, classes of various levels present a dilemma to the support personnel of universities; how is it possible to combine administrative rights for senior students and at the same time "lock" computers for freshmen students? Interestingly enough, commercial applications of hardware and software solutions for virtualized environments provide the resolution to the stated question. Ghostine argues that virtualization of desktops is the answer to the challenges presented to IT support in higher education [6].

Virtualization is based on a time-sharing of hardware concept developed by MIT in 1961 [8]. Virtualization brings server utilization without significant investment, which in its turn boosts return on investment significantly [9]. Interestingly, cloud computing, the basis for virtualization, is also not a novel idea either [10]. Adopting virtualization technologies to teach classes does present challenges but most importantly, opens a wide range of opportunities for faculty to teach newest technologies independent from university IT support. Students' productivity [6] and their interest for Information Technology education increases with the usage of virtualization since full control to fully functional virtual machines (VMs) can be granted without impacting laboratory settings [2]. Besides, via employment of virtualization with server clustering, power consumption, load balancing of servers and the operating cost using x86-based servers is significantly decreased [4], [6].

This manuscript analyzes commercial hardware and software platforms for delivering classes to Information Technology students. Several techniques and approaches including load balancing and power consumption of servers will be discussed.

___

## Server Virtualization

There are common misinterpretations of virtualization technologies. It is important to understand that *virtualization is not emulation or simulation*. In simulation systems, there are mostly software processes that simulate reality in order to attempt to investigate a real process [11]. Emulation, on the other hand, is a process where some original software and hardware environment is being imitated and presented in a different system [12]. Virtualization allows sharing of hardware resources among various entities, and requests are sent and processed through a virtualization layer to underlying hardware. For example, VMware virtualization systems are mostly based on dynamic resource management using tickets. "Resource rights are encapsulated by first-class objects called tickets" [13]. "VMware invented virtualization for the x86 platform in the 1990s to address underutilization and other issues, overcoming many challenges in the process" [14].

Virtualization is now an integral part of commercial IT environments [7], and it is being increasingly utilized with introductions of new technologies, such as vSphere 4.1 and EMC's VPLEX storage systems. There are two main types of virtualization approaches: 1) host-based (Figure 1) and 2) bare-metal hypervisors (Figure 2).

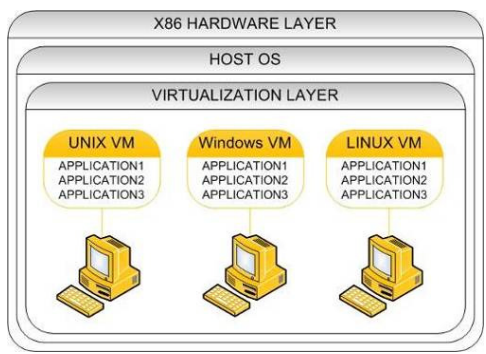

**Figure 1**. **Hosted approach in virtualization [15]**

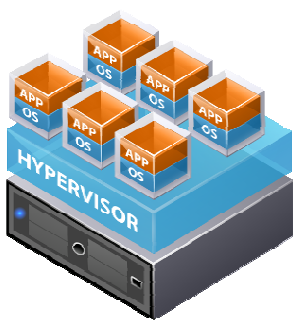

**Figure 2. Hypervisor approach [Source: VMware vSphere 4.1: Install, Configure, Manage – Revision A]**

A host-based approach installs a virtualized layer on top of the existing operating system, Linux or Windows. In the bare-metal approach, *the virtualization layer is the operating system*, which eliminates the OS layer that exists in the host-based approach. Bare-metal hypervisors are the most commonly used in commercial datacenters, whereas host-based virtualization is mostly used in small or educational environments.

Despite the approach to virtualization, many benefits may be enjoyed with either host-based or hypervisor (bare-metal) virtualization [4], [7]. Previously, only commercial systems heavily utilized virtualization to save energy and capital costs via consolidation of physical servers. Today, education facilities use commercial systems mainly for the following reasons [6]: 1) to provide cutting-edge technology for teaching, 2) to create secure, independent classroom environments, and 3) to increase utilization of servers for teaching several classes on the same hardware and software platforms. Additionally, this manuscript describes an innovative approach in energy savings via deployment of Distributed Power Management. Today, the costs of powering and cooling are very significant factors in large-scale datacenters [16].

## Hypervisors

Hypervisors (hosts) are essential hardware and software platforms that create a foundation for virtual infrastructure. Each hypervisor is capable of running multiple various operating systems independent from each other. This approach allows teachers to run multiple unsecure operating systems that do not affect university networks or storage systems. Multiple hypervisors may be managed as clusters by special management software, which eliminates the need for cumbersome administration of each hypervisor separately. There are several clusters of servers that different classes utilize; therefore, the IT1130 cluster would be a cluster of servers for the Introduction to Information Technology course IT1130 and so on. All server clusters are located under 2210 Datacenter, which is the laboratory in the College of Information Technology. For example, Figure 3 depicts an IT1130 cluster of hypervisors that is managed by "VCMAIN08" (hypervisor management platform) vCenter in 2210 datacenter.

## Virtual Machines

A virtual machine (VM) is a set of files that is located on a storage system accessible to hypervisors. Each VM has its folder with a file for memory, BIOS, configuration, virtual disk and others. Figure 4 presents a set of files for a Win-dows 2003 VM. Virtual machines are simply operating systems that are managed by hypervisors.

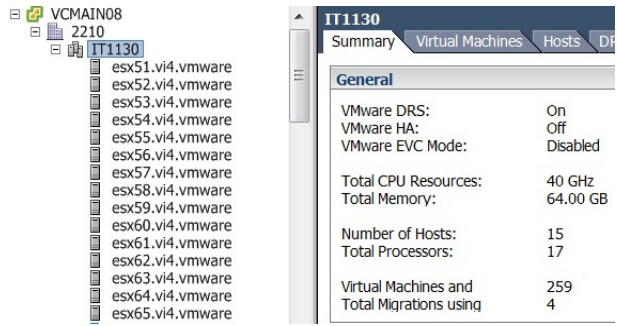

**Figure 3. IT1130 Cluster with 15 ESX Hypervisors**

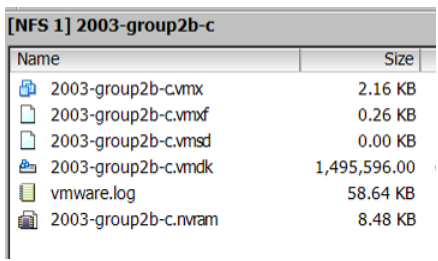

**Figure 4. Files of a Windows 2003 Virtual Computer**

Virtual computers have virtual hardware, BIOS and all other aspects of physical computers. Frequently, the operating system (OS) is not aware that it is running as a virtual machine; it really depends on if the operating system is a virtualization-aware OS.

## Network Storage

Storage is an essential component for virtualized environments [5]. It provides a common storage for VMs, ISOs and other types of files. Depending on the I/O load of storage, commercial-grade systems use different multipathing policies with load balancing on RAID configurations. A number of concurrent server connections can significantly impact the performance of storage systems—specifically, if many students are attempting to deploy multiple VMs on the same storage system simultaneously. Most commonly used for storage systems are Storage Area Network (SAN), Network Attached Storage (NAS) and IP SAN. This manuscript describes several approaches to load balancing of student laboratory operations.

## Teaching Design

The reason virtualization is heavily employed in the learning process of IT students is due to a high level of utilization of hardware of physical servers. In physical, non-virtualized environments, utilization of servers is around 8 – 12%. In virtualized environments, the utilization rates go up to 100%. (Sixty-five percent is the recommended rate.) In education, virtualization presents a faculty member with a powerful environment where multiple technologies and concepts can be learned on the same platform. For example, any type of operating system can be installed in a vSphere 4.1 environment, and multiple networks can be set up for different groups of students. Students in the introductory course learn how to deploy DHCP, DNS, Linux storage systems and many other essential topics. Commercial-grade systems like ESXi servers in a vSphere 4.1 environment allow for a full-scale deployment of IT systems. Free or small environment virtualization products, such as VMware Player, VMware Workstation and others, offer limited functionality in terms of networking, storage and many other options. Figure 5 presents a variety of networks that are used by different students based on their permissions.

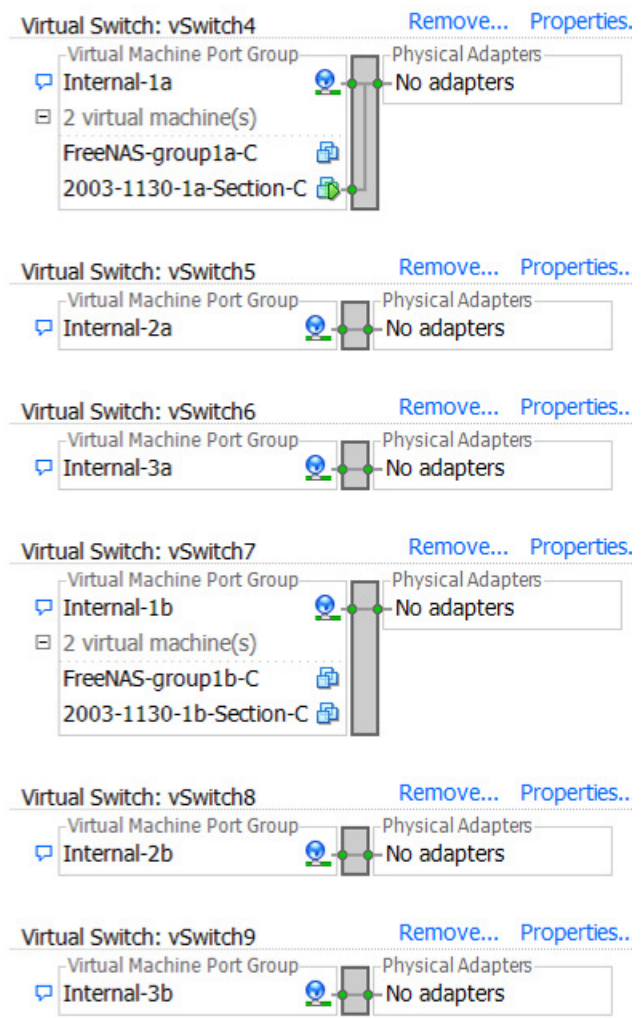

**Figure 5. Multiple Networks for Student of "a" or "b" Groups**

For example, from Figure 5, "Internal 1a, 2a and 3a" networks are designed for students of "a" groups only. Virtual machines "FreeNAS-group1a-C" and "2003-1130-1a-Section-C" ("C" indicates the section of the class) utilize network "Internal 1a." To the students of "b" groups, these networks are not visible. Simply put, groups "a" or "b" of students can *concurrently* deploy DHCP and DNS virtual servers *on the same physical server*, and there are no conflicts or collisions. Such a learning environment for networking is not possible with non-commercial systems; there are no tools, such as Active Directory permissions and privileges (which are used in the presented environment), that can be assigned in free or small environment virtualization products, such as VMware Player, VMware Workstation and other systems.

There are three IT courses that utilize virtualization technologies described by this manuscript: 1) Introduction to Information Technology (freshmen-level), 2) Datacenter Management (senior-level), and 3) Information Storage and Management (senior/graduate-level). The first two courses are described in detail; the third course has originated from an EMC Academic Alliance agreement and is an extensive course on network storage technologies.

Each class is challenged with a number of hands-on laboratory exercises. For example, Introduction to IT has 10 labs, Datacenter Management has 27 labs, and Information Storage and Management has eight labs. The number of hands-on activities has been increasing progressively due to the overwhelming interest of the students who take these courses. Today, IT education is evolving into a form that better accommodates industry needs; furthermore, industries are now eager to sign partnership agreements with educational institutions to better prepare the future workforce [17].

Students' knowledge is assessed through conventional tests and quizzes. Additionally, students submit their labs in the format of an MS Word file with screenshots of accomplished tasks. This approach encourages academic integrity in the classroom and creates a better assessment environment for a faculty member. Windows CIFS network share is used for submissions of the assignments and lab instructions. This network share is located on one of the Openfiler storage servers.

## Introduction to Information Technology Course

There are several labs in the IT1130 course. In fact, the major portion of a student's grade comes from the completion of laboratory exercises, not from tests or quizzes. Each lab has detailed instructions, and there are 10 labs during each semester. For most of the labs, students are required to submit screenshots of completed labs on a designated folder on the shared drive (CIFS share). Students from all classes were successful in sharing one shared drive. The shared drive has helped the faculty member with posting documents, instructions, manuals and specific files for the teaching assistant. Most importantly, the described network share presented the faculty member with the ability to access all of the homework submitted via a centralized location.

At the end of the class, IT1130 students have to decide whether they want to take a hands-on final examination or a regular test.

## Datacenter Management Course

The IT4234 course is a very popular course among IT students. IT4234 is a part of the Networking and Datacenter Administration specialization for IT majors. This class is a three-hour, once-a-week course that presents students with advanced virtualization concepts, and the information presented in this course is the foundation for popular cloud computing technologies. The IT4234 course is 60% hands-on. Unlike the IT1130 courses, there are no screenshots of completed lab assignments required in the IT4234 course. However, the completion of each lab was verified by the instructor and/or teaching assistant. The teaching assistant had taken the same IT4234 course previously and was personally present during each class session. After students complete the course, there is an option to take a certification exam by VMware and become a certified professional. So far, students have positively and enthusiastically commented on the course structure and many became interested in the virtualization subject. If fact, there are several success stories where student received employment offers with Fortune 100 companies.

## Laboratory Infrastructure and Setup

In order to utilize virtualization technology for teaching, it is imperative to have several hardware/software systems in place. The main components of the setup for teaching are: 1) x86 servers, 2) network storage, 3) networking equipment and 4) software and licensing. In order to set up an infrastructure that is independent from the university IT infrastructure and support, a faculty member needs to possess an in-depth knowledge of these systems for correct deployment. Once the systems are in place, ongoing maintenance is required. Maintenance could be achieved without extensive involvement of a faculty member; a teaching assistant with certain rights and privileges to the systems may easily maintain the laboratory environment.

## Physical Servers

There are a total of 31 x86 servers deployed for three courses:

- Fifteen Dell R610 servers are used for Datacenter Management and Information Storage and Management courses.
- Both courses are senior-level with two students receiving access to one server that has 8 GB of RAM, two four-core 2.26 GHz Intel Xeon processors and eight network interface cards (NICs).
- Fifteen IBM x335 servers are used for the Introduction to Information Technology freshmen course, which typically has several sections. At least 60 students utilize a total of 40 GHz of CPU and 64 GB of RAM resources. Each IBM server has at least 4 GB of RAM, one dual-core 2.4 GHz Intel Xeon processor and four network cards.
- One Dell R610 server is used for support and administration of the entire laboratory environment. It has 24 GB of RAM, two four-core 2.26 GHz Intel Xeon processors and eight NICs.

## Shared Storage and Networking

Storage is, perhaps, the most challenging component of the laboratory setup. The reasons are due to the following facts: 1) commercial storage systems are always expensive, and 2) open-source operating systems are a challenge to correctly set up and administer. However, with the use of virtualization technologies, it becomes somewhat easier to administer and deploy different storage environments. The following is a list of storage devices used for the laboratory environment:

- Three Openfiler servers are used for Network File Systems (NFS, used for UNIX/Linux shares). Each server has a dual-core AMD Athlon 7750 processor, 4 GB of RAM and three NICs. One NIC is used for management of the server; the other two are used for load balancing in link aggregation mode. Each server has a 500 GB hard disk for the Openfiler operating system and two 500 GB hard disks in software RAID 0 mode.

- One Openfiler server is used for a Common Interface File System (CIFS, used for Windows network shares), and it has the same configurations of CPU, RAM and network as the other Openfiler servers. There are four 500 GB drives that are set up in a software RAID 0 configuration, for a total of 2 TB of space used for ISO images, support files and, most importantly, assignment submissions from students.

## Resource Utilization

In the laboratory setup, there are several ways that available hardware is used. Thirty-one Dell servers are used for the IT4234 Datacenter Management course, and they are always powered on. Students of the IT4234 course rely on server availability twenty-four hours a day and seven days a week every week since the class requires students to finish 27 laboratory exercises during 16 weeks. The Dell iSCSI server is always on and provides storage for the Dell servers with installed ESX 4.1 operating systems.

Figures 6 and 7 present similar diagrams for both cabinets—IT1130 and IT4234. However, these server cabinets are quite different from each other. The IT4234 cabinet is a powerful pool of resources in comparison to the IT1130 cabinet. The memory of each Dell server is 8 GB, and the instructor's server has 24 GB of RAM. Combined, the total RAM available for the IT4234 course is 264 GB with 560 GHz of CPU cycles. All of these physical resources are not put into a cluster due to the need for each group of students to fully administer each individual server. Students of the IT4234 course have full administration rights to their servers. Currently, there are four servers (including the instructor's server) that are used to support the infrastructure of the IT4234 course.

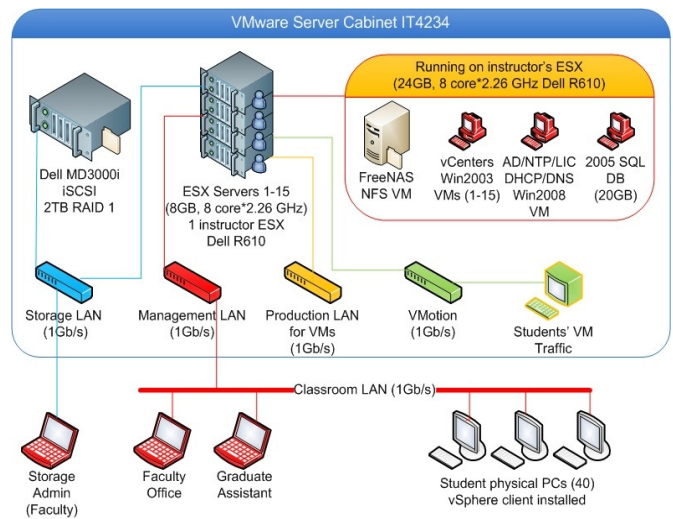

**Figure 6. Infrastructure Setup for the IT4234 Course**

The number of support servers fluctuates since it depends on the number of students that sign up for the class. There are two students assigned to each physical server, so the practical limit of the class size is 15—two students per server, with 30 servers available.

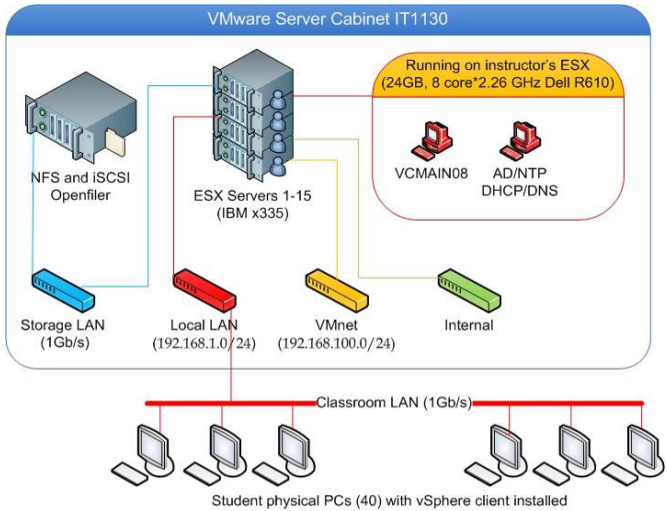

**Figure 7. Infrastructure Setup for the IT1130 Course**

The IT1130 cabinet is extensively used by about 60 students each semester. The resources of this cabinet are utilized in a different way from the utilization of the IT4234 cabinet. IT1130 students have restricted access to the *cluster* (IT1130 cabinet) of resources, not the individual servers. The total resources available in the IT1130 cabinet are 40 GHz of CPU cycles and 64 GB of RAM. Given the fact that the total number of virtual machines is close to 300, the IT1130 cabinet reaches the peaks of its utilization almost every class time.

As mentioned earlier, 15 IBM servers are put into one cluster of servers, which enables load balancing of VMs across the cluster. For example, if a VM machine is seeking resources to utilize, the vCenter will move it as necessary to a different server if no resources are left on the source host. Each student is responsible for each laboratory exercise. Each student has an individual login and credentials, but each section has the same logins. For example, Section C students have logins "group1a, group1b." Section F uses the same logins but creates different VMs. Additionally, the cluster of servers can be put into standby mode to save on power consumption and the cooling of servers.

## Distributed Power Management

IT4234 servers are always powered on and available to students since there is a number of challenging laboratory exercises in the Datacenter Management course. On the other hand, due to the fact that there are 15 physical servers that provide hardware resources to several sections of the IT1130 course, power consumption and cooling of these servers was an important consideration for implementing a green IT solution [18]. Fifteen IBM servers have been set up with power management schema that allowed putting servers in standby mode while servers are not in use. A server schedule was created in order to accommodate students' needs for assignment completions. The schedule was shared with the students of the IT1130 course (both sections C and F). Students are able to come to the laboratory and work on individual assignments. A number of scheduled tasks were set up in order to accommodate server availability. Figure 8 presents one of the scheduled tasks. It shows that the server cluster powers on servers on Monday, Wednesday, Friday, Saturday and Sunday at 11am.

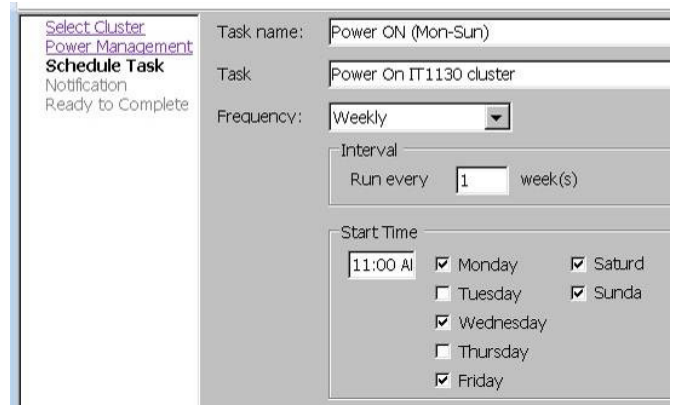

**Figure 8. Scheduled Task to Power a Server Cluster**

A separate task to power the cluster needs to be additionally scheduled. With such implementation of Distributed Power Management (DPM), the laboratory environment becomes power-efficient, and electricity is not wasted. Additionally, when other classes take place, the noise level of servers is significantly reduced once the servers are placed into standby mode. Table 1 provides a sample server schedule for IT1130 servers only; IT4234 servers are always powered on.

**Table 1. Server Distributed Power Management Schedule**

| Time | MWF |
|---|---|
| 8:00 – 8:50a.m. | |
| 9:30 a.m. – 12:15 p.m. | IT 4234A [M] |
| 10:00 – 10:50 a.m. | IT 3234A [WF] |
| 11:00 – 11:55 a.m. | SERVERS ON [MW] |
| 12:00 – 12:50 p.m. | SERVERS ON [MW] |
| 1:00 – 1:50 p.m. | SERVERS ON [MW] |
| 2:00 – 3:15 p.m. | IT 1130C [MW] |
| 3:30 – 4:45 p.m. | IT 1130F [MW] |
| 5:00 – 6:15 p.m. | SERVERS ON [MW] |

Figure 9 depicts servers in standby mode. Two servers are running in case any virtual machines need to power on; additionally, if more hardware resources are required, the cluster-wide settings will wake the servers to provision-requested hardware. With DPM, the server power consumption was dramatically reduced. Such a solution promotes green computing and virtualization technologies, and students realize the benefits of these technologies. Distributed Power Management is mostly implemented in commercial datacenters, and now it is being used in information technology education [18].

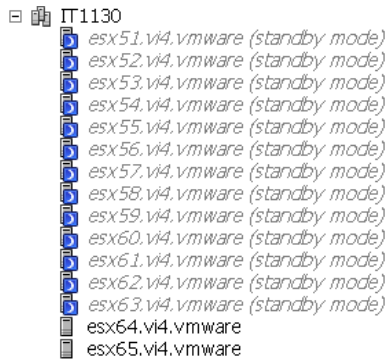

**Figure 9. IT1130 Servers in Standby Mode**

Green computing is a popular terms in today's terminology. By the Distributed Power Management technology presented here, green computing is used for teaching virtualization, storage and other topics in various classes. However, DPM is another feature that is not available in non-commercial virtualization platforms, such as VMware Workstation or VMware Player, that have been reviewed by many educators, such as Dale Lungsford and others [7].

# Challenges and Recommendations

The setup and design of IT courses presented by this manuscript create an opportunity for a faculty member to deliver exciting curriculum with a significant learning experience for students. Students that take these courses receive an opportunity to learn many aspects of IT systems to the full extent. The described virtualization approach does not impact university IT infrastructure in terms of the conventional approach (i.e., without the usage of virtualization).

Based on experience with the one approach to virtualization by one company, students are able to apply their skills in any virtualized environment despite vendor differences. Once the concept of virtualization is understood by students, they do not have any problems adapting to any vendor-specific environments.

One of the biggest challenges in implementing the teaching and the infrastructure setup is the burden on the faculty member's time and his or her involvement with the deployed systems. Once the systems are operational, there should be a teaching assistant with good virtualization and storage skills to take over. Faculty should be prepared to constantly monitor and administer running systems since updates to servers, VMs and configurations may change or require an upgrade.

Based on the experiences resulting from teaching IT courses using virtualization, the following are some recommendations for teaching:

a) Appropriate commercial training on virtualization and network storage is highly recommended for a faculty that is going to deliver courses using virtualized environments. Most of the commercial training can be achieved free of charge with proper agreements with companies that have educational programs. The training requirements are available at the participating industries' websites, and the specific training courses' name are omitted here in order to exclude advertisement in this research publication.

b) A teaching assistant, specifically a student who has taken the Datacenter Management course or who is certified in virtualization technologies, with technical skills is a very important person. Faculty will struggle to deliver the course without any help. The university's IT helpdesk is not expected to participate in preparation for classes, so most of the time a faculty member will rely on the teaching assistant.

c) The teaching assistant may be paid as a student worker for IT services. Most of the students that support virtual infrastructure receive attractive job offers right after graduation.

In addition to teaching recommendations, here are some hardware and software recommendations:

a) Usage of commercial-grade systems is the most desirable setup in laboratories. Most of the described systems are highly available, clustered resources with distributed power management and deployment of resources on demand.

b) Many IT companies are willing to donate equipment to educational IT programs that have a distinct plan for teaching. A big portion of the utilized equipment for the lab environment described by this manuscript was donated free of charge to our IT program.

c) Internal funding, such as technology fee requests and grants, may help to acquire hardware resources. One of the approaches to creating a virtual infrastructure is to deploy so-called Network Development Group (NDG) servers. NDG creates laboratory environments for Cisco and VMware Academies and proved to be a valuable solution in education.

d) Accessibility to the laboratory setup was accomplished through a secured VPN connection to the faculty's office computer, which has two network cards: one for the university network and one for laboratory access. This gave the author access to the lab environment from any PC with Internet access.

Here are some specific recommendations for storage:

a) Openfiler open-source operating systems performed very well with multiple concurrent connections. Originally, FreeNAS open-source system served as NFS share, but Openfiler demonstrated to be a more stable and reliable choice. Openfiler can be deployed on any regular PC and serves as network storage; so the cost of storage can be minimal.

b) NFS performed better than iSCSI for concurrent ESX connections on Openfiler systems.

c) Commercial-grade NICs in link-aggregation (802.3ad) mode is a highly recommended configuration (about $30 each). For example, the use of Intel Gigabit GT network cards on storage servers was found to be very effective and stable with Openfiler systems.

d) Enterprise-level hard disks with at least 32 MB of onboard cache memory are recommended. Desktop/home-grade hard disks are not recommended. Seagate Constellation ES ST31000524NS 1 TB drives performed very dependably under multiple I/O simultaneous requests from all the servers.

e) The described storage systems worked great with software-based Linux RAID configuration (default software RAID configuration in Openfiler), so there was no need to purchase expensive RAID controllers.

f) It is recommended to distribute the load of server connections to storage across multiple storage servers. The way storage servers distributed the load of I/O processing was in the following fashion; there were three NFS Openfiler systems. Each one served five groups of students, and one Openfiler server served as a CIFS share for the submission of all homework and for information sharing.

g) The cost of Dell servers was about $3,000 – $4,000 for each student server and $5,000 – $6,000 for the instructor server. A Dell iSCSI server was purchased for about $13,000; however, all of these prices have changed dramatically, and many open-source systems today can replace expensive hardware.

Virtualization has greatly enhanced the learning experience for our IT students. Additionally, our university received international recognition for teaching advanced curriculum.

# Biography

**TIMUR MIRZOEV** is a professor of Information Technology at Georgia Southern University, College of Information Technology. Dr. Mirzoev heads the International VMware IT Academy Center and the EMC Academic Alliance programs at Georgia Southern University. Some of Timur's research interests include server and storage virtualization, cloud computing, storage networks and disaster recovery. Currently, Dr. Mirzoev holds the following certifications: VMware-certified instructor (VCI), VMware-Certified Professional (VCP) 4, EMC- Proven Professional, LefthandNetworks (HP), SAN/iQ and A+. Dr. Mirzoev may be reached at tmirzoev@georgiasouthern.edu